# Decoherence and Collisional Frequency Shifts of Trapped Bosons and Fermions

Kurt Gibble

Department of Physics, The Pennsylvania State University, University Park, PA 16802
LNE-SYRTE, Observatoire de Paris, 75014 Paris, France

We perform exact calculations of collisional frequency shifts for several fermions or bosons using a singlet and triplet basis for pairs of particles. The "factor of 2 controversy" for bosons becomes clear – the factor is always 2. Decoherence is described by singlet states and they are unaffected by spatially uniform clock fields. Spatial variations are critical, especially for fermions which were previously thought to be immune to collision shifts. The spatial variations lead to decoherence and a novel frequency shift that is not proportional to the partial density of internal states.

Spectroscopy is a powerful tool for the study of ultracold gases. Examples include probing their phases and excitations [1-4], precise atom-interferometers [5,6], and accurate atomic clocks, ranging from atomic fountains [7-10] and optical frequency lattice clocks [11-15] to chip-scale clocks [16]. In these, collisions of particles produce frequency shifts that often limit performance [7-9,16,17]. There has been confusion and controversy about the frequency shifts of systems of bosons and fermions as they lose coherence [1,15,17-23]. Here we present a unified picture of the frequency shifts of bosons and fermions with exact calculations of systems of several weakly interacting particles. The picture, based on the coherent evolution of singlet and triplet combinations of pairs of particles, clearly describes the role of decoherence in the spectroscopy of ultracold systems.

For bosons, there is a long-standing puzzle, the "factor of 2 controversy" [1,17-23]. It suggests that frequency shifts should change as decoherence allows identical bosons to be distinguished. The factor of 2 describes the doubling of the scattering rate, due to exchange symmetry, when two identical bosons collide. If identical bosons are prepared in different internal states, there is no exchange symmetry so there is no factor of 2. More interestingly, when both bosons are prepared in identical superpositions of two internal states, there is exchange symmetry and the scattering and frequency shifts are enhanced by a factor of 2. But, in the presence of interstate decoherence, there is no longer exchange symmetry of these superpositions and therefore the interstate scattering cannot be enhanced by a factor of 2. Therefore it seems that the frequency shift must change as Bose gases decohere. However, no change was observed – "the Ramsey Fringe that Didn't Chirp" [17]. This led to the unsettling deduction that the factor of 2 does not change as gases decohere. Our picture shows that identical superpositions do indeed have a factor of 2 enhancements. But, in the presence of decoherence, there are pair-wise pseudo-spin singlet states of particles and these are unaffected by a spatially homogeneous clock field. Therefore, although particles in the singlet states are distinguishable, decoherence does not affect the frequency shift because the homogeneous clock field only couples to triplet states which describe identical coherent superpositions. Further, we show that the spatial variation of the clock field mimics decoherence and produces a novel frequency shift due to interstate decoherence, analogous to the elusive chirp [17].

It has been widely believed that trapped identical fermions are immune to collisional frequency shifts [20,23] at ultracold temperatures where there is only s-wave scattering. But recently, a collisional shift of an optical frequency fermion clock was observed [15]. We show that the spatial variation of the clock field, naturally larger for optical fields as compared to radio-frequency fields treated previously [20-23], produces the same novel shift due to interstate decoherence. We also show that fermion collision shifts can be suppressed, allowing highly-accurate future optical frequency clocks.

We begin with a Hamiltonian for N bosons or fermions:

$$H = H_0 + \tfrac{1}{2}\sum_{\substack{i=1\ldots N \\ \eta=a,b,c\ldots}} \Omega_\eta \left|S\psi_\eta(\vec{r}_i)\right\rangle\left\langle P\psi_\eta(\vec{r}_i)\right|$$

$$+ \Omega_\eta^* \left|P\psi_\eta(\vec{r}_i)\right\rangle\left\langle S\psi_\eta(\vec{r}_i)\right| + \sum_{i<j}^{N} V(\vec{r}_i - \vec{r}_j)$$

Here $H_0$ describes the trap degrees of freedom (Fig. 1a). We take $H_0$ to be independent of the internal states S and P, as in an optical lattice clock [11] and strong enough to be in either the resolved sideband [11-15] or Lamb-Dicke limits [16,17,23] so that Doppler shifts and photon recoils can be neglected. The wavefunction for the system is comprised of products of single particle wavefunctions that are products of spatial wavefunctions, $\psi_a(\mathbf{r_1})$, $\psi_b(\mathbf{r_2})$…, and internal state wavefunction S and P, where a,b,c… denote trap states and 1, 2, 3… are particle labels. We allow a general coupling of the clock field to each trap state, $\Omega_a$, $\Omega_b$, $\Omega_c$ … with a form $\Omega_a = \Omega_{0a} e^{i\Delta\omega_a t + i\phi_a}$ in a rotating frame where $\Delta\omega_a$ is the detuning, $\phi_a$ is the phase, and similarly for $\Omega_b$, $\Omega_c$…. The spatial amplitude and phase variations of the clock field yield couplings that depend on the trap state [15, 24]. Fig. 1b depicts two particles, in external trap states a and b, Rabi flopping between the two internal states with a small difference in couplings. The interaction of the particles is described by V($\mathbf{r_i}$–$\mathbf{r_j}$). We treat interactions that are weak compared to the trap frequencies and neglect higher order corrections from scattering between trap states.

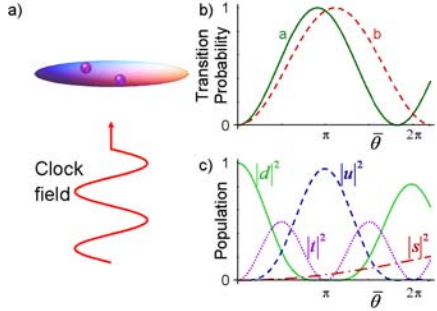

Fig. 1. (Color online) a) Schematic of spectroscopy of trapped particles. b) Transition probability of particles in trap states *a* and *b* as a function of clock pulse area $\bar{\theta}$. The coupling of the clock field to the particles in state *a* is larger than for *b*. c) The populations of two-particle triplet and singlet states corresponding to b). The small difference of Rabi frequencies leads to a small amplitude of the singlet state s.

The nature of collision shifts for many particles emerges for just 2 particles in a trap. A natural basis for two identical particles is the pseudo-spin singlet and triplet states of the internal states:

$$\begin{aligned}|d\rangle &= |S_1 S_2\rangle \psi_{ab}^\pm(\vec{r}_1,\vec{r}_2) \\ |t\rangle &= 2^{-1/2}|S_1 P_2 + P_1 S_2\rangle \psi_{ab}^\pm(\vec{r}_1,\vec{r}_2) \\ |s\rangle &= 2^{-1/2}|S_1 P_2 - P_1 S_2\rangle \psi_{ab}^\mp(\vec{r}_1,\vec{r}_2) \\ |u\rangle &= |P_1 P_2\rangle \psi_{ab}^\pm(\vec{r}_1,\vec{r}_2)\end{aligned} \quad (1)$$

Here $|d\rangle$, $|t\rangle$, and $|u\rangle$ are triplet states, $|s\rangle$ is a singlet state, and the spatial wave function $\psi_{ab}^\pm(\mathbf{r_1,r_2})=2^{-1/2}$ [$\psi_a(\mathbf{r_1})\psi_b(\mathbf{r_2})\pm\psi_b(\mathbf{r_1})\psi_a(\mathbf{r_2})$] is symmetric (+) or antisymmetric (−). The upper (lower) sign in Eq. 1 is for bosons (fermions). Only symmetric states have s-wave interactions – the triplet states for bosons and the singlet for fermions.

A general state for two particles is $u|u\rangle+t|t\rangle+s|s\rangle+d|d\rangle$. We first discuss fermions and then return to bosons below. The equations of motion for two fermions are:

$$i\dot{d} = \tfrac{\bar\Omega}{\sqrt2}t - \tfrac{\Delta\Omega}{\sqrt2}s \quad i\dot{t} = \tfrac{\bar\Omega}{\sqrt2}u + \tfrac{\Omega^*}{\sqrt2}d \\ i\dot{u} = \tfrac{\bar\Omega^*}{\sqrt2}t + \tfrac{\Delta\Omega^*}{\sqrt2}s \quad i\dot{s} = \tfrac{\Delta\Omega}{\sqrt2}u - \tfrac{\Delta\Omega^*}{\sqrt2}d + 2g_{SP}s \quad (2)$$

The average and difference of Rabi frequencies are $\bar{\Omega}=(\Omega_a+\Omega_b)/2$ and $\Delta\Omega=(\Omega_a-\Omega_b)/2$. The fermion interactions are given by $g_{SP}$. To lowest order, $g_{SP}$ is $(2\hbar a_{SP}/m)\int|\psi_{ab}^+(\mathbf{r,r})|^2 dV$ where $a_{SP}$ is the scattering length. In Fig. 1c, $\bar{\Omega}$ couples $|t\rangle$ to $|d\rangle$ and $|u\rangle$ whereas the singlet state $|s\rangle$ is weakly coupled to $|d\rangle$ and $|u\rangle$ by $\Delta\Omega$.

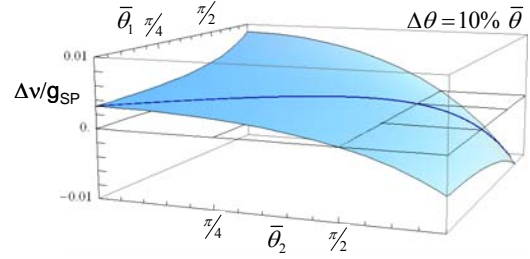

Fig. 2. (Color online) Collisional frequency shift of fermions as a function of the Ramsey pulse tipping angles, $\bar{\theta}_1$ and $\bar{\theta}_2$. The shift is not proportional to the difference of partial densities $n_P$–$n_S$, which would imply a strong dependence $\bar{\theta}_1$. The shift does vary strongly with the second Ramsey pulse area, going to 0 if the second pulse is on average a $\pi/2$ pulse. The solid line is the shift for the usual case of identical pulses.

We use Eq. 2 to calculate the frequency shift for Ramsey spectroscopy [7-10]. Two clock pulses, separated by a long interrogation time, are applied to particles prepared in S [25]. We calculate the transition probability, $2|u|^2+|t|^2+|s|^2$, as a function of frequency. The collision shift is:

$$\Delta\nu = \frac{g_{SP}}{\pi A}\sin(2\Delta\theta_1)\sin(\Delta\theta_2)\cos(\bar{\theta}_2) \quad (3)$$

Here, the tipping angles are $\bar{\theta}_j = \bar{\Omega}\tau_j$ and $\Delta\theta_j = \Delta\Omega\tau_j$, where $\tau_j$ is the pulse length, j=1 or 2 refers to the first or second Ramsey pulse, and the amplitude of the Ramsey fringes is $A = \sum_{k=a,b}\sin(\theta_{1k})\sin(\theta_{2k})$, approximately $N\sin(\bar{\theta}_1)\sin(\bar{\theta}_2)$ for N particles and small $\Delta\Omega$.

The frequency shift is 0 as expected if the two fermions have the same Rabi frequencies for the first Ramsey pulse ($\Delta\theta_1$=0) and are therefore indistinguishable [20-23]. If there is a difference of Rabi frequencies, the first pulse creates a singlet amplitude, s=$2^{-1/2}$ i sin($\Delta\theta_1$). During the interrogation time, the singlet state acquires a phase shift due to $g_{SP}$ and gives a frequency shift proportional to $g_{SP}\sin(\Delta\theta_1)$. In Fig. 2, we show the collisional shift versus the tipping angles $\bar{\theta}_1$ and $\bar{\theta}_2$. Unlike in homogenous gases, the shift is not proportional to the difference of the partial densities, $n_P$–$n_S$ – it is nearly independent of $\bar{\theta}_1$ in Fig. 2. For weak pulses, the shift is nearly constant. The shift strongly depends on the second pulse's area, $\bar{\theta}_2$. For $\bar{\theta}_2=\pi/2$, the shift goes to zero even if the first pulse creates a singlet amplitude that produces a large mean field energy. Below we show that the two particles have large and opposite frequency shifts and these exactly cancel for $\bar{\theta}_2=\pi/2$. Further, if the second Ramsey pulse is homogeneous ($\Delta\theta_2$=0), there is no frequency shift in Eq. 3 because the second pulse does not couple to the singlet state so the transition probability $2|u|^2+|t|^2+|s|^2$ is unaffected by a phase shift of s. Often, the second Ramsey pulse is the same as the first. The solid lines in Figs. 2 and 3 show $\Delta\nu$ for identical pulses, $\bar{\theta}_1=\bar{\theta}_2$. For many fermions, the equations of motion can be successively rotated so that each pair-wise interaction is pure singlet or triplet. The shift is the sum of Eq. 3 for all pairs, $\Delta\nu=(N-1)g_{avg}\Delta\theta^2\cos(\bar{\theta})/\pi\sin^2(\bar{\theta})$, where $\Delta\theta$ is small, $\Delta\theta^2$ is the rms spread of tipping angles, and $g_{avg}$ is the average of the interactions $g_{SP}$.

Coherence and decoherence are at the core of the nature of ultracold gases. Zwierlein et al. observed that fermions prepared in coherent superpositions of internal states, even after they fully decohered, exhibited no frequency shift [23]. This is noteworthy because different phases of the superpositions make the fermions distinguishable and they therefore interact. They considered a uniform clock field and showed that fermions are immune to frequency shifts. A key point was that the clock pulse reintroduces coherence and changes the two-particle spatial correlation function $g^{(2)}$ [23]. They used the quantum optics definition of $g^{(2)}$, $\langle n_P n_S\rangle/\langle n_P\rangle\langle n_S\rangle$, which describes the collision probability per ground and per excited state atom, not the collision probability per atom pair $G^{(2)}=\langle n_P n_S\rangle/n^2$ [19]. In our picture, dephasing splits a population, initially prepared in $|t\rangle$, between the singlet state $|s\rangle$ and triplet state $|t\rangle$. For triplet states, $G^{(2)}$ and $g^{(2)}$ are both 0 whereas, for a singlet state, $G^{(2)}=1/2$ and $g^{(2)}=2$. Since a homogeneous clock field does not couple to the singlet state, there is neither a change of $G^{(2)}$ nor a frequency shift. Therefore, $g^{(2)}$ in [23] changes not because the number of collisions change, but because the clock field transfers atoms from $n_P$ to $n_S$, changing the normalization of $g^{(2)}$. More generally, a clock field does have a spatial variation, which changes $G^{(2)}$ and the number of collisions, giving fermions a collision shift. The shift is proportional to the singlet amplitude but also depends critically on the evolution during the second Ramsey pulse.

Recently, a density dependent frequency shift for fermions was observed in a lattice clock [15]. Many lattice clocks currently use a single Rabi pulse instead of two Ramsey pulses [11-15]. As in Fig. 3, the frequency shift for a single Rabi pulse (dashed) behaves nearly identically to that for two Ramsey pulses (solid) [26]. Ref. 15 used $g^{(2)}$ to argue that the frequency shift is proportional to the difference of internal state densities, $n_P - n_S$. Although their model fits their data, their data also fits our model (Fig. 3 inset). We arrive at the different conclusion that the collision shift is generally not 0 when $n_P=n_S$, near $\Delta\omega=0.76\ \bar{\Omega}$ in the inset of Fig. 3. We note that their 1μK data is also consistent with no collision shift with an excess $\chi^2$ of 0.9. For 3μK, our model fits with a $\chi^2$ of 4.1 for 6 degrees of freedom but here, higher lattice vibrational states are excited and therefore scattering induced tunneling between lattice sites could give a different frequency shift that would depend strongly on the trap depth and be non-linear in density. The variation of the frequency shift with trap depth and its linearity with density are straightforward to test experimentally, as is the suppression of the shift by varying the pulse area.

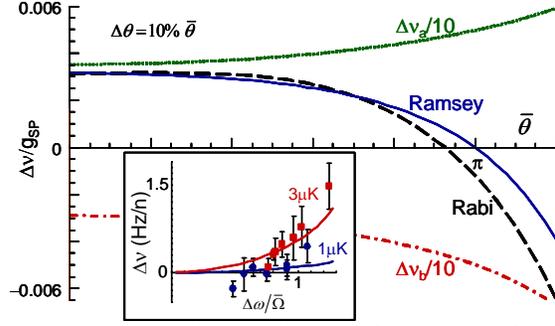

Fig. 3. (Color online) Collisional frequency shifts of fermions versus the total clock pulse area $\bar{\theta}$. The two Ramsey pulses (solid) are identical $\bar{\theta}=\bar{\theta}_1+\bar{\theta}_2$ and the shift for a single Rabi pulse (dashed) behaves similarly. The Rabi detuning is the half width for a π pulse. Two colliding fermions have large and nearly opposite shifts that nearly cancel, $\Delta\nu_a$ (dotted) and $\Delta\nu_b$ (dash-dot), reduced by 10 times. The inset shows the data of [15] for a Rabi π pulse versus detuning $\Delta\omega$ and our model. Neither cross 0 as a function of detuning, as occurs with pulse area $\bar{\theta}$. We infer detunings from measured transition probabilities.

It is instructive to compare interactions in a trap to scattering in free space. If two fermions in arbitrary superpositions collide [10], the scattering yields an outgoing s-wave with a singlet amplitude $\sin(\Delta\theta_1)$. The interference of the scattered s-wave, which is an entangled superposition of S and P, with the unscattered part of each fermion's wavefunction produces the frequency shift. The interference of S with one fermion implies the interference of P with the other, and vice versa, giving opposite frequency shifts to each fermion. In Fig. 3, the dotted curve shows a large frequency shift $\Delta\nu_a$ for one fermion, nearly independent of $\bar{\theta}$, and a nearly opposite shift, $\Delta\nu_b$, for the other. Thus, suppressing fermion collision shifts requires homogeneous excitation [24], control of the pulse area, and reasonably uniform detection.

Bosons are also susceptible to the frequency shift due to interstate decoherence. The equations of motion are the same as Eqs. 2 but with particle interactions $2(g_{SS}, g_{SP}, 0, g_{PP})$ for (d, t, s, u), where $g_{XY}$ has the appropriate scattering length, $a_{SS}$, $a_{PP}$, or $a_{SP}$. Subtracting an energy of $2\bar{g}\equiv g_{PP}+g_{SS}$ from all states, gives

$$i\dot{d}=\tfrac{\bar{\Omega}}{\sqrt{2}}t-\tfrac{\Delta\Omega}{\sqrt{2}}s-\Delta g d$$
$$i\dot{u}=\tfrac{\bar{\Omega}^*}{\sqrt{2}}t+\tfrac{\Delta\Omega^*}{\sqrt{2}}s+\Delta g u$$
$$i\dot{t}=\tfrac{\bar{\Omega}}{\sqrt{2}}u+\tfrac{\bar{\Omega}^*}{\sqrt{2}}d+\delta g t$$
$$i\dot{s}=\tfrac{\Delta\Omega}{\sqrt{2}}u-\tfrac{\Delta\Omega^*}{\sqrt{2}}d-2\bar{g}s$$

Here $\Delta g\equiv g_{PP}-g_{SS}$ and $\delta g\equiv 2g_{SP}-g_{PP}-g_{SS}$. Therefore, when all scattering lengths are equal ($a_{SS}=a_{PP}=a_{SP}$), $\Delta g=\delta g=0$ and the bosons behave identically to fermions, but with the opposite frequency shift [27]. In general, the frequency shift for thermal bosons and Ramsey pulses is:

$$\Delta\nu = \frac{\Delta g}{2\pi} + \frac{\delta g}{\pi}\cos(\bar{\theta}_1)\cos(\Delta\theta_2)\frac{\sin(\bar{\theta}_1)\sin(\bar{\theta}_2)}{A}$$
$$-\frac{\bar{g}}{\pi A}\sin(2\Delta\theta_1)\sin(\Delta\theta_2)\cos(\bar{\theta}_2)\quad(4)$$

The first two terms are well known for homogeneous gases [17,28]. The first is the difference of the mean field energy due to the intrastate scattering. The second term is proportional to the partial density, $n_S-n_P = n\cos(\bar{\theta}_1)$, and the triplet amplitude t after the first pulse, $\sin(\bar{\theta}_1)$. The third term is the same as Eq. 3 and describes the singlet state – this population is absent from the triplet state and doesn't get a shift from $g_{SP}$ to cancel that from $-2\bar{g}$.

The "factor of 2 controversy" for bosons concerns the "2" in $\delta g \equiv 2g_{SP}-g_{PP}-g_{SS}$ [1,17 -23]. If there is interstate decoherence, it seems that the factor of 2 should decrease to 1 [17]. A thesis of Ref. 17 was that interstate decoherence could be probed by varying the partial density $n_P-n_S$ because the second term in Eq. 4 is proportional to $\delta g$ and $n_P-n_S$. Here, this term *always* has the factor of 2, regardless of interstate decoherence. Decoherence can only change the collision shift via the singlet amplitude and it is unaffected by homogeneous clock fields. In this picture, interstate and intrastate density correlations were in fact different in [17]. Just as for fermions, if the second Ramsey pulse is not a $\pi/2$ pulse and its inhomogeneity is correlated with that of the singlet state, the large and elusive frequency shift, or chirp, can be observed as bosons decohere.

In summary, clocks based on fermions are not immune to collisional frequency shifts. The spatial variation of the clock field makes otherwise identical fermions distinguishable by directly exciting pair-wise singlet states that have s-wave collisions. The collisions lead to a novel frequency shift of the clock that is proportional to the density, but not the partial density $n_P-n_S$. The frequency shift is generally not the difference of mean field energies of the initial and final states but depends on the coherent evolution of the particles in the clock field. By controlling the strength of the second clock pulse in a Ramsey sequence, the shift can be precisely suppressed if the particles are detected reasonably uniformly. For a 1D fermion optical lattice clock, the scale of the interactions is $g_{SP}/2\pi \approx 1$ to 30 Hz for $a_{SP} \approx 300\ a_0$. If the rms variation of the clock coupling is 10% [24], and the clock pulse area is controlled to 1%, the frequency shift is of order 2 mHz, or fractionally $10^{-17}$ to $10^{-18}$ [29]. The picture of pair-wise singlet and triplet interactions also makes the "factor of 2 controversy" for Bose gases clear. The factor of 2 is always 2 because only particles in identical superpositions participate in the transition. The loss of interstate coherence can lead to a large frequency shift if the clock field is spatially inhomogeneous. It is the same as the shift for fermions, but with the opposite sign.

We acknowledge assistance from S. Kokkelmans and D. Petrov, conversations with J. Banavar, S. Bize, Z. Hadzibabic, W. Ketterle, F. Laloe, P. Lemonde, K. O'Hara, E. Tiesinga, P. Rosenbusch, J. Thomsen, Ch. Salomon, B. J. Verhaar, D. Weiss, J. Ye, and M. Zwierlein, and financial support from NASA, NSF, ONR, and la ville de Paris.

Since this work was completed, others have treated fermion collision shifts as proportional to the difference of mean field energies and partial densities [30-32]. Our model gives different predictions, e.g. Eq. 3.